\documentclass[aps,prb,twocolumn,showpacs]{revtex4-1}
\usepackage{amsfonts}
\usepackage{amsmath}
\usepackage{amsthm}
\usepackage{graphics}
\usepackage{graphicx}
\usepackage{subfigure}
\usepackage{amssymb}
\usepackage{graphics}
\usepackage{graphicx}
\usepackage{epstopdf}
\usepackage{color}
\usepackage{subfigure}
\usepackage{pgfplots,tikz}
\usepackage{url}
\usepackage{float}
\usepackage{soul}
\newcommand\bet{{g}}
 
\newcommand\alps{{\frac{\hbar^2}{2m}}}

\newcommand\dertt[1]{ \frac{\partial{ #1}}{\partial t} }
\newcommand\gd{\mbox{${\bf \nabla}^{2}$}}
\newcommand\grad{\mbox{${\bf \nabla}$}}

\def\Mp{M_{\rm I}}
\def\Rp{a_{\rm I}}
\def\vp{\mathbf{\dot{q}}}

\def\Vp{V_\mathrm{I}}
\def\x{\mathbf{x}}
\def\q{\mathbf{q}}
\def\p{\mathbf{p}}

\def\k{\mathbf{k}}
\def\kmax{k_\mathrm{max}}
\def\tauB{\tau_{\mathrm{I}}}
\def\tauGP{\tau_{\mathrm{GP}}}

\begin{document}

\title{Stochastic motion of finite-size immiscible impurities in a dilute quantum fluid at finite temperature}
\author{Umberto Giuriato}
\affiliation{
Universit\'e C\^ote d'Azur, Observatoire de la C\^ote d'Azur, CNRS, Laboratoire Lagrange, Nice, France}
\author{Giorgio Krstulovic}
\affiliation{
Universit\'e C\^ote d'Azur, Observatoire de la C\^ote d'Azur, CNRS, Laboratoire Lagrange, Nice, France}

\begin{abstract}
The dynamics of an active, finite-size and immiscible impurity in a dilute quantum fluid at finite temperature is characterized by means 
of numerical simulations of the Fourier truncated Gross--Pitaevskii equation. The impurity is modeled as a localized repulsive potential 
and described with classical degrees of freedom. It is shown that impurities of different sizes thermalize with the fluid and undergo a stochastic dynamics compatible with an Ornstein--Uhlenbeck process at sufficiently large time-lags. The velocity correlation function and the displacement of the impurity are measured and an increment of the friction with temperature is observed. Such behavior is phenomenologically explained in a scenario where the impurity exchanges momentum with a dilute gas of thermal excitations, experiencing an Epstein drag. 
\end{abstract}
\maketitle

\section{Introduction}

A Bose--Einstein  condensate (BEC) is an exotic state of matter, which takes place in bosonic systems below a critical temperature, 
when a macroscopic fraction of particles occupy the same fundamental quantum state \cite{pitaevskii2016bose}. Almost three decades ago, 
Bose--Einstein condensation was observed for the first time by Anderson et al. in a dilute ultra-cold atomic gas \cite{BECExp95}. 
Since then, BECs have been realized in a wide range of different systems, from solid-state quasiparticles \cite{kasprzak2006bose, demokritov2006bose} to light in optical micro-cavities \cite{klaers2010bose}.

Bose--Einstein condensation is intimately related to the notion of superfluidity, which is the capability of a system to flow without viscous dissipation \cite{pitaevskii2016bose}. Superfluidity was first detected almost one century ago in liquid helium 
$^4\mathrm{He}$ \cite{Kapitza1938,Allen1938} below 2.17K, and it is a known feature also of atomic BECs and light in nonlinear optical systems \cite{CarusottoSuperfluidsLight}. Both superfluidity and Bose--Einstein condensation are a manifestation of quantum effects on a macroscopic scale, which is why these systems are usually called quantum fluids. Theoretically, a quantum fluid can be described by a macroscopic complex wave function. This represents the order parameter of the Bose--Einstein condensation phase transition and it is directly related to the density and the inviscid velocity of the superflow via a Madelung transformation \cite{noretal}. 

As a consequence of superfluidity, an impurity immersed in a quantum fluid does not experience any drag and can move without resistance. However, if the speed of the impurity is too large, superfluidity is broken because of the emission of topological defects of the order parameter, known as quantum vortices \cite{donnelly1991quantized,FrischCritVel,BrachetCritVel,ActiveWiniecki}. Moreover, at finite temperature the thermal excitations in the system may interact with the impurities and drive their motion \cite{ClusteringUmberto}. The behavior of particles and impurities immersed in a superfluid has been a central subject of study since long time \cite{donnelly1991quantized}. The interest has been recently renewed by the experimental implementation of solidified hydrogen particles to visualize quantum vortices in superfluid helium \cite{bewley2006superfluid, LaMantiaParticles}, the study of polarons in atomic gases \cite{Impurity_BEC,Tracers_BEC} and the use of impurities to investigate the properties of superfluids of light \cite{MichelSuperLight,CarusottoLight2014}. A particularly interesting kind of impurity arises in the immiscible regime of the multi-component BEC. It has been shown that when two condensates of different species highly repel each other, one of the two components exists in a localized region and can be thought as a finite-size impurity \cite{NonlinearBEC_book,RicaRoberts}. If many components are present simultaneously, different phases can be identified, depending on the ratios between the coupling constants \cite{RicaRoberts}. In particular, for {positive scattering lengths} between the impurity fields, the components separate from the main condensate and show a hard-sphere repulsion between each other. Experimentally, mixtures of different condensates have been realized with cold atomic gases \cite{Modugno2components,Myatt2components}, and the immiscibility properties have been studied \cite{Papp2components}.

In this work we aim at studying numerically the dynamics of an immiscible and finite-size impurity in a quantum fluid at finite temperature. There are several models which have been proposed to take into account finite temperature effects in a quantum fluid, although at the moment there is no uniform consensus on which is the best one \cite{ProukakisFiniteTemperature}. A successful example is the Zaremba-Nikuni-Griffin framework, in which a modified-dissipative Gross--Piteaevskii equation for the condensate wavefunction is coupled with a Boltzmann equation for the thermal cloud \cite{ZNG}. A simpler model is the Fourier truncated Gross--Pitaevskii (FTGP) equation, in which thermal fluctuations of the bosonic field are naturally taken into account without the coupling with an external thermal bath \cite{DavisFiniteTEmpBEC}. The main idea behind the FTGP model is that imposing an ultraviolet cutoff $k_{\mathrm{max}}$, and truncating the system in Fourier space, allows for the regularization of the classical ultraviolet divergence and states at thermal equilibrium can be generated. The FTGP model has been successfully used to reproduce the condensation transition \cite{DavisFiniteTEmpBEC,DavideBKT,CondensationRica,KrstulovicBottleneck}, to study finite temperature effects on quantum vortex dynamics \cite{BerloffRing,KrstulovicSlowdown,GiorgioFiniteTempPRE} and to investigate the effective viscosity in the system \cite{ShuklaViscosity}.

In this article, we couple the FTGP equation with a minimal model for impurities, which are described as localized repulsive potentials with classical degrees of freedom \cite{ActiveWiniecki,ShuklaParticlesPRA2017}. It has been recently utilized systematically to investigate the interaction between particles and quantum vortices at very low temperature \cite{GiuriatoApproach,GiuriatoKelvinwaves,GiuriatoReconnections,GiuriatoTangle}. We stress that this minimal model is suitable for extensive numerical simulations and Monte-Carlo sampling. Indeed, its simplicity makes it computationally much cheaper than more complex approaches in which the impurities have many (infinite) degrees of freedom, like the Gross--Clark model \cite{BerloffBubble,VilloisBubble} or the multi-component BEC model \cite{RicaRoberts}.

Recently, a drag force acting on an impurity in the weak coupling regime has been detected using a damped GP equation at finite temperature \cite{SpanishDrag}, extending an analytical work in which the resistance of the GP fluid on a point particle was studied at zero temperature \cite{PitaevskiiTheory}. In the case of immiscible active impurities, it has been shown that a multitude of them coupled with the FTGP model can form clusters, depending on the temperature and the ratio between the fluid mediated attraction and the impurity-impurity repulsion \cite{ClusteringUmberto}. Moreover, the presence of such clusters turned out to be responsible for an increase of the condensation temperature. However, the precise characterization of the dynamics of a single impurity immersed in a bath of FTGP thermal modes has not been addressed yet. This is indeed the purpose of the present work. In the next section, we present the FTGP model coupled with a single three-dimensional impurity, and provide details for the numerical techniques used to simulate such system. In section \ref{sec:impurity_motion}, we present a statistical analysis of extensive numerical simulations of the system. In particular, we find that at large times the dynamics of an impurity in a finite temperature quantum fluid is akin to an Ornstein--Uhlenbeck process with a temperature dependent friction coefficient, that we are able to explain. Eventually, we exploit this information to show that for the sizes of the impurities considered, their motion is consistent with a scenario where the thermal excitations behave as a gas of waves rather than a continuum liquid. 

\section{Finite temperature model}
\label{sec:model}
We use the Fourier truncated Gross-Pitaevskii model to describe a weakly interacting quantum fluid at finite temperature, with a repulsive impurity immersed in it \cite{ClusteringUmberto}. The Hamiltonian of the model is given by:
\begin{eqnarray}
H&=&\int\left( \frac{\hbar^2}{2m} |\grad \psi |^2 +\frac{g}{2}|\mathcal{P}_{\rm G}[|\psi|^2]|^2\right)\,\mathrm{d} \x  + \nonumber \\
&&\int\Vp(| \x - \q |)\mathcal{P}_{\rm G}[|\psi|^2]\,\mathrm{d} \x +\frac{\p^2}{2 \Mp} ,
\label{Eq:HGP}
\end{eqnarray}
where $\psi(\x,t)$ is the bosonic field, $m$ is the mass of the constituting bosons and $g=4 \pi  a_\mathrm{s} \hbar^2 /m $ is the self-interaction coupling constant, with $a_{\rm s}$ the bosons $s$-wave scattering length.

The bosonic field is coupled with an impurity of mass $\Mp$, described by its classical position $\q(t)$ and momentum $\p(t)=\Mp\mathbf{\dot{q}}(t)$. The impurity is modeled by a repulsive potential $\Vp(|\x-\q|)$, which defines a spherical 
region centered in $\q(t)$ where the condensate is completely depleted.
Note that the functional shape of the potential $\Vp(|\x-\q|)$ is not important, provided that it is sufficiently repulsive to completely deplete the fluid. The relevant parameter is indeed the size of the depleted region, which in turns identifies the impurity radius $\Rp$. The Galerkin projector $\mathcal{P}_{\rm G}$ truncates the system imposing an UV cutoff in Fourier space: $\mathcal{P}_{\rm G} [\hat{\psi}_{\k}] = \theta(\kmax-|\k|)\hat{\psi}_{\k}$ 
with $\theta(\cdot)$ the Heaviside theta function, $\hat{\psi}_\k$ the Fourier transform of $\psi(\x)$ and $\k$ the wave vector.
The time evolution equation of the wavefunction and the impurity are obtained straightforwardly by varying the Hamiltonian (\ref{Eq:HGP}): 
\begin{equation}
i\hbar\dertt{\psi}=\mathcal{P}_{\rm G} \left[- \alps \gd \psi + \bet\mathcal{P}_{\rm G} [|\psi|^2]\psi+  \Vp(| \x -{\bf q}|)\psi\right], \label{Eq:GPE}
\end{equation}
\begin{equation}
\Mp\frac{\rm d \mathbf{\dot{q}}}{\rm d t}= - \int  \Vp(| \x -{\bf q} |)  \mathcal{P}_{\rm G}[\nabla|\psi|^2]\, \mathrm{d} \x. \label{Eq:Particles}
\end{equation}
Note that the projection of the density $|\psi|^2$ in Eq.\eqref{Eq:GPE} is a de-aliasing step that is necessary to conserve momentum \cite{GiorgioFiniteTempPRE} in the truncated equations. This procedure slightly differs with the Projected Gross--Pitaevskii model \cite{DavisFiniteTEmpBEC} as some high-momentum scattering processes are not considered in the FTGP framework.

At zero temperature and without the impurity, Eq.(\ref{Eq:GPE}) can be linearized about the condensate ground state $\psi_0=|\psi_0|\exp{(-i\mu t/\hbar)}$, fixed by the chemical potential $\mu=g|\psi_0|^2$. The excitations of the condensate propagate with the Bogoliubov dispersion relation:
\begin{equation}
\omega_\mathrm{B}(k) = ck\sqrt{1+\frac{\xi^2k^2}{2}},
\label{Eq:bogo}
\end{equation}
where $k=|\k|$, 
$c=\sqrt{g|\psi_0|^2/m}$ is the speed of sound and $\xi=\sqrt{\hbar/{2gm|\psi_0|^2}}$ defines the healing length at zero temperature. Note that the impurity completely depletes the condensate in the region where $V_{\mathrm{I}}>\mu$.

The Hamiltonian $H$ and the number of bosons $N=\int |\psi|^2\mathrm{d} \x$ are invariants of the FTGP model. 
Thus, it possesses finite temperature absolute equilibrium solutions, distributed with the probability
\begin{equation}
\mathbb{P}[\psi,\mathbf{q},\mathbf{\dot{q}}]\propto e^{-\beta(H - \mu N)}. 
\label{Eq:equilibrium}
\end{equation}
The concept of absolute equilibria of Fourier truncated equations was first introduced in the context of the Euler equation \cite{Lee1952,Kraichnan1967} and directly generalizes to FTGP \cite{GiorgioFiniteTempPRE}. Such equilibria are steady solutions of the associated Liouville equation. The Liouville equation describes the microcanonical evolution of the phase-space distribution function of an ensemble of states driven by Eqs. (\ref{Eq:GPE},\ref{Eq:Particles}).
Note that a state which solves Eqs. (\ref{Eq:GPE},\ref{Eq:Particles}) conserves the invariants $N$ and $H$, and the equilibrium distribution in Eq. \eqref{Eq:equilibrium} is nothing but the probability of picking one of these states at given inverse temperature $\beta$ and chemical potential $\mu$.
This is true whether the impurity is present in the system or not. The argument of the exponential in Eq. (\ref{Eq:equilibrium}) is a linear combination of the invariants $H$ and $N$, and $\beta$ is a Lagrange multiplier identified with the inverse temperature. Given a random initial condition with energy $H$ and number of bosons $N$, long time integration of the equations (\ref{Eq:GPE},\ref{Eq:Particles}) will let the system evolve to an equilibrium state belonging to the distribution (\ref{Eq:equilibrium}). The temperature is not directly available as a control parameter, since such dynamics is microcanonical, but it is biunivocally associated to the given conserved invariants \cite{DavisFiniteTEmpBEC}. 

At finite temperature, many modes are excited and interact non-linearly. Such interactions lead to a spectral broadening of the dispersion relation, together with small corrections of the frequency.  Overall, the dispersion relation can be well approximated taking into account the depletion 
of the condensate mode in the following manner  \cite{ShuklaViscosity}:
\begin{equation}
\omega^T_\mathrm{B}(k) = ck\sqrt{n_0(T)+\frac{\xi^2k^2}{2}},
\label{Eq:bogoT}
\end{equation}
where $n_0(T)$ is the condensate fraction. We define it as 
\begin{equation}
n_0(T) = \frac{\left\langle |\int\psi\,\mathrm{d}\x|^2 \right\rangle_T}{\left\langle |\int\psi\,\mathrm{d}\x|^2 \right\rangle_{T=0}}, 
\label{Eq:condensate}
\end{equation}
namely as the ratio between the occupation number of the zero mode at temperature $T$ and at temperature $T=0$. With such definition, the condensate fraction is normalized to be one at zero temperature. In this way, the depletion of the condensate due to the presence of the impurity is properly taken into account \cite{ClusteringUmberto}. The fraction of superfluid component $n_\mathrm{s}(T)=\rho_\mathrm{s}/\bar{\rho}$ and normal fluid component $n_\mathrm{n}(T)=\rho_\mathrm{n}/\bar{\rho}$, 
where $\bar{\rho} = \frac{1}{L^3}\int m|\psi|^2\,\mathrm{d}\x$ is the average mass density, can be computed using a linear response approach \cite{ClarkDerrikSuper,FosterBKT,ClusteringUmberto}. They read, respectively:
\begin{equation}
n_\mathrm{n}(T)=\frac{\lim_{k\rightarrow 0}\chi_I(k)}{\lim_{k\rightarrow 0}\chi_C(k)},\qquad\quad 
n_\mathrm{s}(T) = 1 - n_\mathrm{n}(T),
\label{Eq:chiratio}
\end{equation}
where $\chi_C(k)$ and $\chi_I(k)$ are respectively the compressible (longitudinal) and incompressible (transverse) coefficients 
of the 2-points momentum correlator:
\begin{equation}
\left\langle \hat{j}_i(\mathbf{k})\hat{j}_j(\mathbf{-k}) \right\rangle \propto\frac{k_ik_j}{k^2}\chi_C(k)+\left(\delta_{ij}-\frac{k_ik_j}{k^2}\right)\chi_I(k),
\label{Eq:momcorr}
\end{equation}
with $\hat{j}_i(\mathbf{k},t)$ the Fourier transform of the $i$-th component of the momentum density $j_i(\mathbf{x},t)=\frac{i\hbar}{2}\left[\psi\partial_i\psi^*-\psi^*\partial_i\psi\right]$.

\subsection*{Numerical methods and parameters}

In the numerics presented in this work, we integrate the system (\ref{Eq:GPE},\ref{Eq:Particles}) by using a pseudo-spectral method with $N_{\mathrm{res}}=128$ uniform grid points per direction of a cubic domain of size $L=2\pi$. We further set the UV cutoff $\kmax=N_{\mathrm{res}}/3$, so that, besides the Hamiltonian $H$ and the number of bosons $N$, the truncated system (\ref{Eq:GPE},\ref{Eq:Particles}) conserves the total momentum ${\bf P}=\int \frac{i\hbar}{2}\left( \psi {\bf \nabla}\psi^* - \psi^* {\bf \nabla}\psi\right)\mathrm{d} \x+\p$ as well (provided that initially $\mathcal{P}_{\rm G} [\psi]=\psi$ and $\mathcal{P}_{\rm G} [\Vp]=\Vp$) \cite{GiorgioFiniteTempPRE,GiuriatoReconnections}. {In thermal states, the cutoff $\kmax$ plays an important role. The dimensionless parameter $\xi\kmax$ controls the amount of dispersion of the system and therefore the strength of the non-linear interactions of the BEC gas. The smaller its value, strongest the interaction is. Note that, as scales of the order of the healing length have to be resolved numerically, it cannot be arbitrarily small. See for instance references \cite{GiorgioFiniteTempPRE,ShuklaViscosity} for further discussions}. In this work we fix this parameter to $\xi k_\mathrm{max}=2\pi/3$. Note that in our results all the lengths are expressed in units of the healing length at zero temperature $\xi$ and the velocities in units of the speed of sound $c$ at zero temperature. {In this units, the system size is $L=128\xi$}.

The potential used to model the impurity is a smoothed hat-function $\Vp(r)=\frac{V_0}{2}(1-\tanh\left[\frac{r^2 -\eta_a^2}{4\Delta_a^2}\right])$. The impurity radius $\Rp$ is estimated at zero temperature by measuring the volume of the displaced fluid 
$\frac{4}{3}\pi \Rp^3 = \int (|\psi_0|^2 - |\psi_\mathrm{p}|^2)\,\mathrm{d}\mathbf{x}$, where $\psi_\mathrm{p}$ is the steady state with one impurity. The impurity mass density is then $\rho_\mathrm{I} = \Mp/\left(\frac{4}{3}\pi \Rp^3\right)$. In all the simulations we fix $\mu=|\psi_0|=1$ and for the impurity potential $V_0=20\mu$ and $\Delta_a = 2.5\xi$. We consider an impurity of radius $\Rp=7.6\xi$ setting $\eta_a = 2\xi$ and an impurity of size $\Rp=12.7\xi$ setting $\eta_a = 10\xi$. 

\begin{figure}
    \includegraphics[width=.99\linewidth]{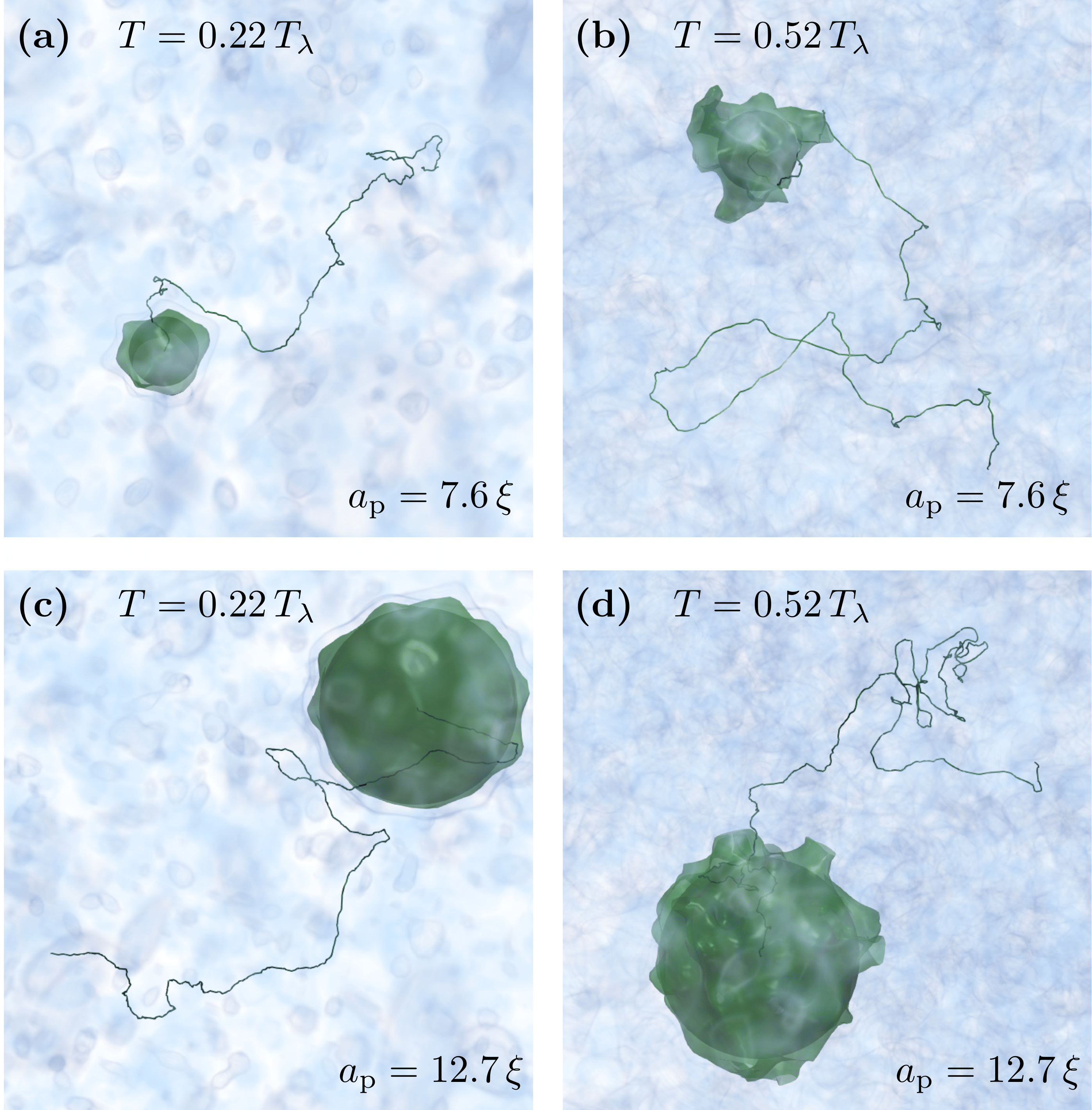}
    \caption{
        (\textit{Color online}) Snapshots of the GP field 
    with an impurity of size $\Rp=7.6\xi$ at time $t=3056\xi/c$ \textbf{(a,b)}
    and an impurity of size $\Rp=12.7\xi$ at time $t=7130\xi/c$ \textbf{(c,d)}
    at temperatures $T=0.22\,T_\lambda$ \textbf{(a,c)} 
    and $T=0.52\,T_\lambda$ \textbf{(b,d)}. 
    The GP sound waves are rendered in blue, the dark sphere is the 
    impurity potential and the green surfaces are contours of the 
    GP density at $\rho/\bar{\rho}=0.15$.
    The impurity trajectory is displayed as a solid line. }
    \label{Fig:3Dtraj}
\end{figure}

Note that, although the shape of the impurity potential is fixed, fluctuations of the impurity surface are allowed by the model. 
Such fluctuations are shown in Fig.\ref{Fig:3Dtraj} (that will be commented in Section \ref{sec:impurity_motion}) as green contours of the fluid density at a low value around the spherical potential.

We prepare separately the ground state with an impurity $\psi_\mathrm{p}$ (at zero temperature) and the FTGP states at finite temperature $\psi_T$, without the impurity. The first one is obtained by performing the imaginary time evolution of the equation (\ref{Eq:GPE}), while the second one is realized with the stochastic real Ginzburg--Landau (SRGL) \cite{GiorgioFiniteTempPRE,ClusteringUmberto,ShuklaViscosity}, protocol that allows to explicitly control the temperature. The SRGL method is briefly recalled below. The initial condition for the FTGP simulations is then obtained as $\psi = \psi_\mathrm{p}\times\psi_T$. For our analysis, {we considered $\sim 22$ different realizations for each of the $15$ studied temperatures and for each impurity}. The initial velocity of the impurity is always set to zero and the temporal length of each realization is $\sim 9000\,\xi/c$. In all the statistical analysis presented in the following sections, we checked that including or not the data associated to the early times of the simulation does not change the results. The thermalization of the impurity will be studied explicitly in the next Section \ref{sec:impurity_motion}, but this fact gives already a first indication that the impurity reaches the equilibrium with the thermal bath in the very early stages of the simulations.

We operatively define the condensation temperature $T_\lambda$ as the first point of the temperature scan at which the condensate fraction $n_0(T)$ goes to zero. The normal fluid fraction $n_\mathrm{n}(T)$ and consequently the superfluid fraction $n_\mathrm{s}(T)=1-n_\mathrm{n}(T)$ are evaluated numerically with the following protocol \cite{FosterBKT}. 
At fixed temperature, we measure the angle--averaged incompressible and compressible spectra of the momentum correlator,  respectively $\chi^{1d}_I(k)\propto\left\langle k^2|\mathbf{j}_I(\mathbf{k})|^2 \right\rangle$ and $\chi^{1d}_C(k)\propto\left\langle k^2|\mathbf{j}_C(\mathbf{k})|^2 \right\rangle$. We fit the logarithm of $\chi^\mathrm{1d}_I(k)/k^2$ and $\chi^\mathrm{1d}_C(k)/k^2$ with a cubic polynomial in the range $3\cdot L/2\pi < k < 3k_\mathrm{max}/2$; we extrapolate the values of the fits at $k=0$ and finally divide them to get $n_\mathrm{n}(T)=\chi_I(k=0)/\chi_C(k=0)$. Such method works well at low temperatures while it is strongly affected by numerical noise at temperatures $T\gtrsim T_\lambda$ \cite{FosterBKT}. These last points are then simply assumed to be equal to zero.

Finally note that in this work, if not explicitly specified, all the averages are intended over realizations for a fixed temperature $T$. Moreover, because of isotropy, we treat each dimension of any vectorial quantity as a different realization of the same distribution.

\subsection*{Grand-canonical thermal states}

We recall here the SRGL protocol used to obtain equilibrium thermal states of the truncated GP equation. 
We refer to Ref.\cite{GiorgioFiniteTempPRE} for further details about the method. The FTGP grand-canonical thermal states 
obey the (steady) Gibbs distribution which coincides with Eq. (\ref{Eq:equilibrium}). A stochastic process that converges to a realization of this probability distribution is given by the following stochastic equation (in physical space):
\begin{eqnarray}
\hbar\frac{\partial \psi}{\partial t} &=& \mathcal{P}_{\rm G} [\alps \gd \psi + \mu\psi - \bet\mathcal{P}_{\rm G} [|\psi|^2]\psi+  \Vp(| \x -{\bf q}|)\psi] \nonumber \\
&&+ \sqrt{\frac{2\hbar}{\beta L^3}} \mathcal{P}_{\rm G} [\zeta(\x,t)],
\label{Eq:stoch}
\end{eqnarray}
where $\zeta(\x,t)$ is a complex Gaussian white noise with zero mean and delta-correlated in space and time: 
$\left\langle \zeta (\x,t) \zeta^* (\x',t') \right\rangle = \delta(\x - \x') \delta(t-t')$. In principle such process is coupled with analogous equations for the impurity degrees of freedom \cite{ClusteringUmberto}. Here, we do not consider them, since we are interested in generating thermal states without impurities. As explained in the previous section, the impurity is added 
afterwards to the thermal states in order to observe its dynamics according to the evolution equations (\ref{Eq:GPE},\ref{Eq:Particles}). 
In the right hand side of Eq. \eqref{Eq:stoch} a deterministic term and a stochastic term compete against each other. The distribution which entails the balance between such fluctuations and dissipation is Eq. \eqref{Eq:equilibrium}, i.e. the steady solution of the Fokker--Planck equation associated to Eq. \eqref{Eq:stoch} \cite{GiorgioFiniteTempPRE}.

We define the temperature as $T=1/k_\mathcal{N}\beta$, where $k_\mathcal{N} = L^3/\mathcal{N}$ and $\mathcal{N} = \frac{4}{3}\pi k_\mathrm{max}^3$ is the number of Fourier modes in the system. With this choice, the temperature has units of energy density 
and the intensive quantities remain constant in the thermodynamic limit, that is $k_\mathrm{max}\rightarrow \infty$ 
with $L$ constant. Finally, in order to control the steady value of the average density $\bar{\rho}$, the chemical potential is also dynamically evolved with the \emph{ad hoc} equation $\dot{\mu} = -\nu_\rho(\bar{\rho}-\bar{\rho}_\mathrm{t})$ during the stochastic relaxation. In this way, the system converges to the control density $ \bar{\rho}=\bar{\rho}_\mathrm{t}$ that we set equal to $m|\psi_0|^2=1$.

We finally mention that a similar approach can be used to generate and study thermal states, which is the stochastic GP model \cite{ProukakisFiniteTemperature}. There, the stochastic relaxation (\ref{Eq:stoch}) is combined with the physical GP evolution (\ref{Eq:GPE}). However, unlike the FTGP model, the stochastic GP model is dissipative and has an adjustable parameter in which the interaction between the condensate and the thermal cloud is encoded.

\section{Impurity motion}
\label{sec:impurity_motion}
We perform a series of numerical simulations of the model (\ref{Eq:GPE},\ref{Eq:Particles}), varying the temperature and the size of the impurity. Typical impurity trajectories are displayed in Fig.\ref{Fig:3Dtraj} for two different temperatures, together with a volume rendering of the field and of the impurity. The motion of the impurity is clearly driven by a random force, due to the interaction with the thermal excitations of the condensate. 

\begin{figure}
\includegraphics[width=.99\linewidth]{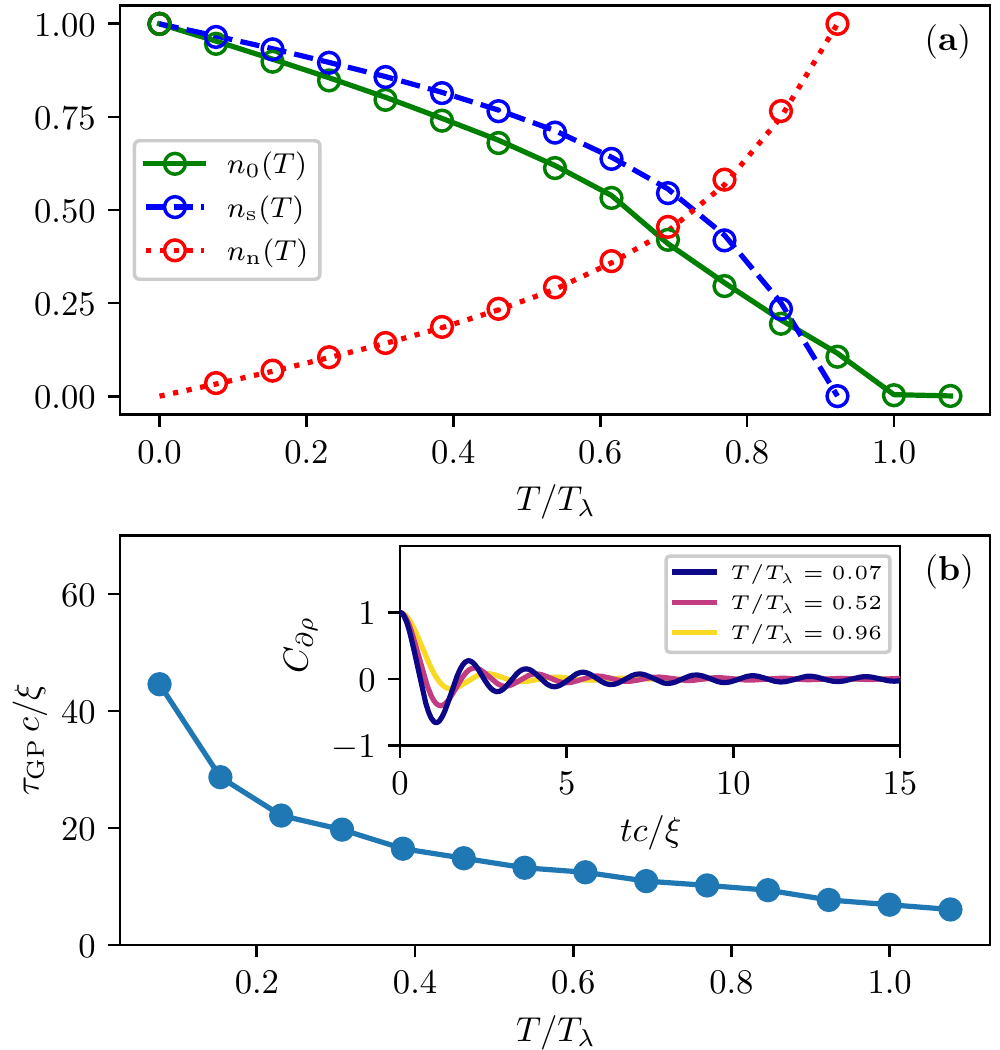}
\caption{
(\textit{Color online}) Temperature evolution of condensate fraction 
(green solid line), superfluid fraction (dashed blue line) 
and normal fraction (dotted red line) for simulations 
without impurity. 
The circles of corresponding colors refer to simulations 
in presence of an impurity  of size $\Rp=12.7\xi$ 
and mass density $\rho_\mathrm{I}=\bar{\rho}$. 
\textbf{(b)} Temperature evolution of the decorrelation time of the FTGP density gradients.
(\textit{inset}) Time evolution of the two-points correlators of the FTGP density gradients 
(\ref{Eq:DensGradCorr}) 
for three different temperatures.}
\label{Fig:Cond}
\end{figure}

Before studying the stochastic dynamics of the impurity, we characterize some properties of the thermal states that will be used later. In Fig.\ref{Fig:Cond}.a we show the condensate fraction $n_0$, the superfluid component $n_\mathrm{s}$ and the normal fluid component $n_\mathrm{n}$ plotted against temperature. 
The lines refer to the simulations without impurity while the circles are obtained in presence of the largest impurity considered ($\Rp=12.7\xi$). Almost no difference between the two cases is detected, since the volume occupied by the impurity is only $0.5\%$. Indeed, in Ref. \cite{ClusteringUmberto} it was shown that the condensate fraction starts to 
increase at high temperatures if the impurities filling fraction is larger than $4\%$. We can therefore safely assume that the impurity has no impact on the statistical properties of the thermal fluctuations.

From the impurity Eq.~\eqref{Eq:Particles}, we observe that the quantum fluid interacts with the impurity via a convolution between the impurity potential and the density gradient. It is thus interesting to understand the typical correlation time of density fluctuations, in particular of its gradients. 
In Fig.\ref{Fig:Cond}.b we compute the decorrelation time $\tauGP$ of the thermal excitations as a function of temperature. 
Such time is evaluated performing a FTGP evolution of thermal states without impurity and considering 
the time correlator of one of the component of the density gradient:
\begin{equation}
    C_{\partial\rho}(t) = \frac{ \left\langle \partial_i\rho(t_0)\partial_i\rho(t_0+t) \right\rangle 
   } { \left\langle (\partial_i\rho)^2 \right\rangle}.
\label{Eq:DensGradCorr}
\end{equation}
The averages in Eq. (\ref{Eq:DensGradCorr}) are performed over space and different realizations. Three examples for three different temperatures of the time evolution of this correlator are shown in the inset of Fig.\ref{Fig:Cond}.b. They show a damped oscillating behavior and touch zero for the first time after a time $\sim 1 c/\xi$. We estimate the decorrelation time $\tauGP$ as the time after which the correlator (\ref{Eq:DensGradCorr}) is always less than $1\%$. At timescales larger than $\tauGP$, we expect that the interactions between the impurity and the thermal excitations can be considered as random and rapid.
Before checking if this is the case, we verify explicitly whether the impurity reaches the thermal equilibrium with the quantum fluid.

\begin{figure}
\includegraphics[width=.99\linewidth]{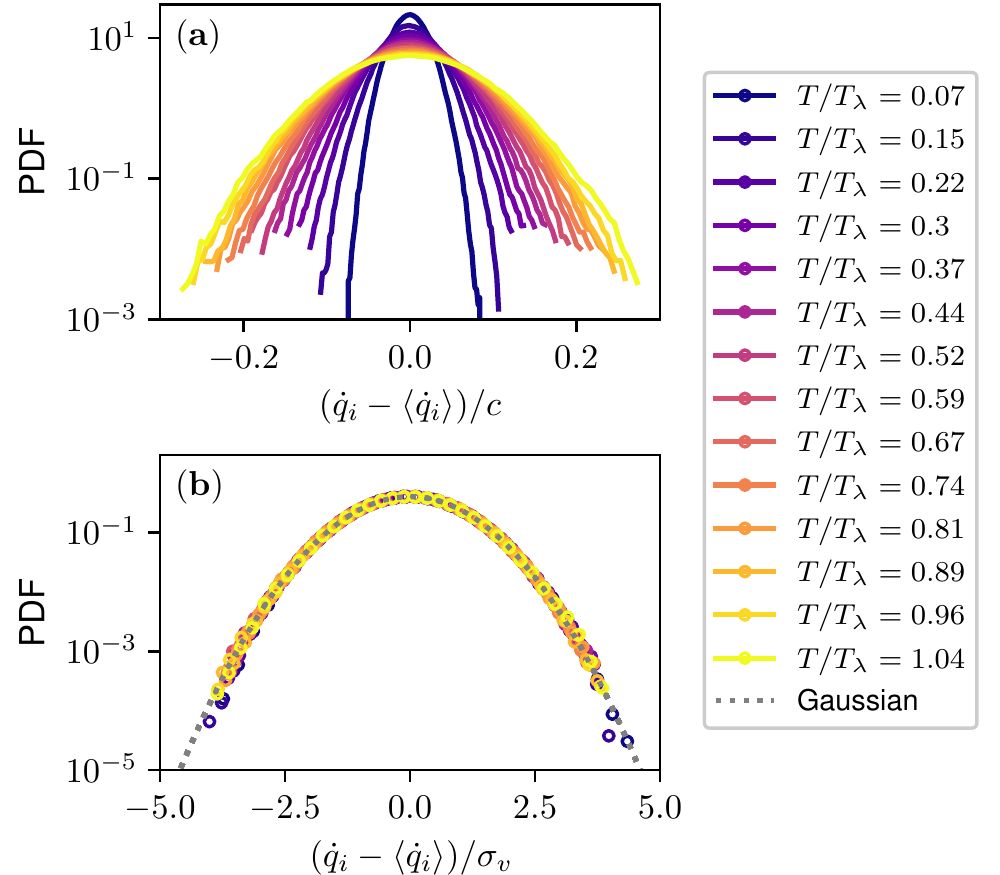}
\caption{
(\textit{Color online}) PDF of the single component velocity 
of an impurity of size $\Rp=7.6\xi$ and mass density $\rho_\mathrm{I}=\bar{\rho}$, for different temperatures. 
\textbf{(a)} Velocities normalized with the 
speed of sound at zero temperature. 
\textbf{(b)} Velocities normalized with the standard deviation. 
Dotted black line is a Gaussian distribution with zero mean and unit variance.}
\label{Fig:PDF}
\end{figure}

\begin{figure}
\includegraphics[width=.99\linewidth]{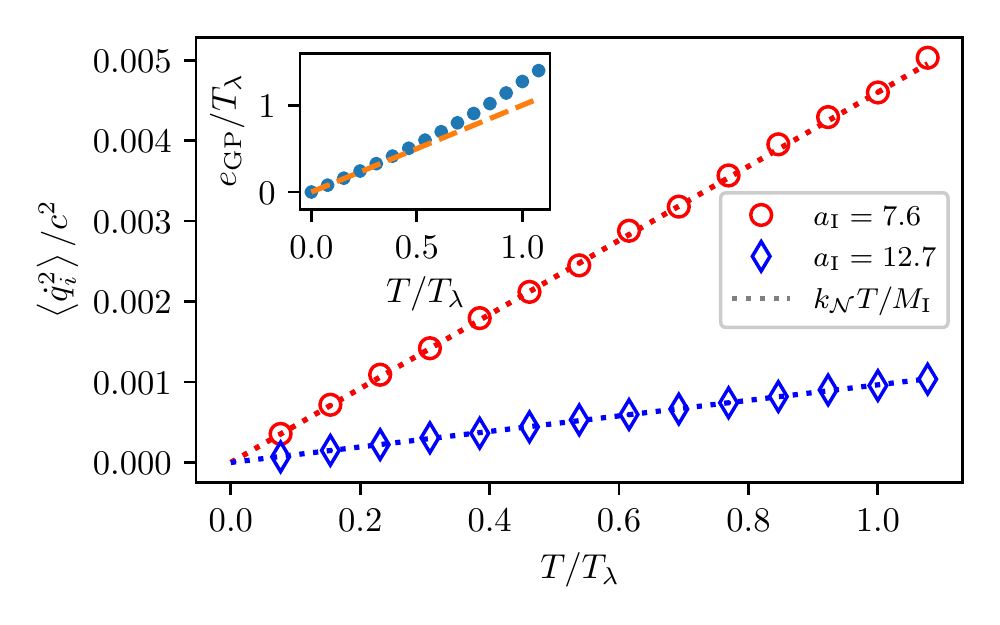}
\caption{
(\textit{Color online}) Second order moment of the single component 
velocity of impurities of size $\Rp=7.6\xi$ (red circles) 
and $\Rp=12.7\xi$ (blue diamonds), 
as a function of the temperature. 
The mass density is $\rho_\mathrm{I}=\bar{\rho}$ for both.
(\textit{inset}) GP energy density versus temperature 
(blue points). Orange dashed line is the equipartition line 
$e_\mathrm{GP}=T_\lambda$.}
\label{Fig:KBT}
\end{figure}

If the number of the excitations-impurity interactions is large, the velocity of the impurity is expected to be normally distributed at the equilibrium, in accordance with the central limit theorem. Indeed, we show this in Fig.\ref{Fig:PDF}, 
where the probability density function (PDF) for the single component of the impurity velocity is displayed. 
Assuming ergodicity, the PDFs are computed averaging also over time, besides over realizations. Since we expect the impurity to be in thermal equilibrium with the surrounding GP fluid, the second order moment of its velocity should relax to a constant value, that is related to the temperature via the equipartition of energy:
\begin{equation}
\left\langle \dot{q}^2_i \right\rangle = \frac{k_\mathcal{N} T}{\Mp}.
\label{Eq:variance}
\end{equation} 
The perfect agreement between Eq.~(\ref{Eq:variance}) and the numerical simulations is displayed in Fig.\ref{Fig:KBT}.
It confirms that the impurity is indeed in thermal equilibrium with the thermal bath.
Note that the linear scaling with temperature persists also at high temperatures, where the GP energies are not in equipartition anymore because of high nonlinear interactions. This is not a contradiction, since the impurity is a classical object with 
a simple quadratic kinetic energy. For comparison, the deviation from equipartition of the GP energy density {$e_\mathrm{GP} = (H-\mu N)/L^3+\mu^2/2g$} (without impurities) is reported in the inset of Fig.\ref{Fig:KBT}.

\begin{figure}
\includegraphics[width=.99\linewidth]{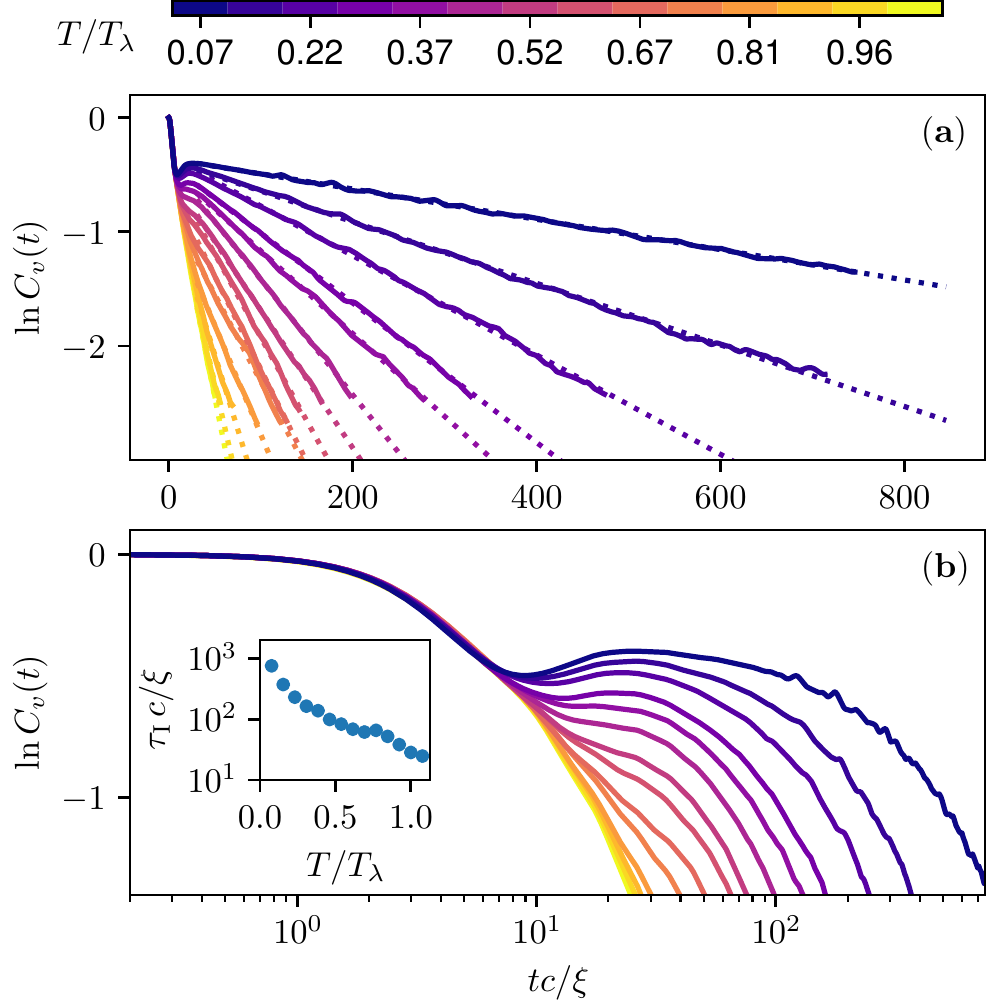}
\caption{
(\textit{Color online}) Time evolution of the two-points velocity correlator for 
the impurity of size $\Rp=7.6\xi$ 
and mass density $\rho_\mathrm{I}=\bar{\rho}$
in \textbf{(a)} Log-Lin scale and \textbf{(b)} Log-Log scale. 
Different colors are associated to different temperatures 
(same legend of Fig.\ref{Fig:PDF}). 
Dotted lines are linear fits. 
(\textit{inset}) Temperature evolution of the dynamical correlation time of the impurity.
}
\label{Fig:Corr}
\end{figure}

We consider now the evolution of the two-point impurity velocity correlator $C_v(t)$. If the collisions between the superfluid thermal excitations and the impurity are fast and random, we expect it to decay as 
\begin{equation}
    C_v(t) = \lim_{t\rightarrow\infty} \frac{ \left\langle \dot{q}_i(t_0)\dot{q}_i(t_0+t) \right\rangle - \left\langle \dot{q}_i \right\rangle^2 } { \left\langle \dot{q}_i^2 \right\rangle - \left\langle\dot{q}_i\right\rangle^2 } = e^{-\frac{t}{\tauB}}.
    \label{Eq:correlator}
\end{equation}
where $\tauB$ is the dynamical correlation time of the impurity velocity.
Specifically, the behavior (\ref{Eq:correlator}) should certainly hold at time-lags larger than the 
decorrelation time of the GP excitations $\tauGP$, estimated in Fig.\ref{Fig:Cond}.b.
This scenario is confirmed by the measurements of $C_v(t)$, reported in Fig.\ref{Fig:Corr} 
for the impurity of size $\Rp=7.6\xi$. 
The exponential decay is evident for time-lags larger than $\sim 10\xi/c$ for all the temperatures.

According to the results mentioned so far, at sufficiently large timescales the interactions between the impurity and the thermal bath can be considered to be effectively fast, random and decorrelated.
Thus, it is natural to suppose that the impurity dynamics may be described by the 
Ornstein-Uhlenbeck (OU) process \cite{VanKampen}: 
\begin{equation}
    \Mp\mathbf{\ddot{q}}= -\gamma\vp + \sqrt{\sigma^2}\mathbf{\zeta}_\mathrm{r}(t),
    \label{Eq:OU}
\end{equation}
where $\mathbf{\zeta}_\mathrm{r}(t)$ is a (Gaussian) white noise in 
time, i.e. 
$\left\langle \mathbf{\zeta}_\mathrm{r}(t) \right\rangle = 0$ 
and 
$\left\langle \mathbf{\zeta}_{\mathrm{r},i}(t_1)\mathbf{\zeta}_{\mathrm{r},j}(t_2) \right\rangle = \delta_{ij} \delta(t_1-t_2)$  
where $\sigma^2$ is related to the diffusion coefficient. 
The term $-\gamma\mathbf{\dot{q}}$ is the drag force, with $\gamma$ a friction coefficient that in general may depend on temperature and on the impurity size. In particular, the friction should be directly related to exponential decay timescale $\tauB$ of the correlator (\ref{Eq:correlator}) as $\gamma=\Mp/\tauB$. In Fig.\ref{Fig:Corr} we clearly see that the correlators decay faster for higher temperatures. The values of the correlation time $\tauB$ at different temperatures are obtained through linear fits of $\ln C_{v}(t)$, shown as dotted lines in Fig.\ref{Fig:Corr}.a. The decreasing of $\tauB$ with temperature is then explicitly displayed in the inset of Fig.\ref{Fig:Corr}.b. Note that $\tauB \gg\tauGP$, consistently with the assumputions of the OU process.
The physical consequence of such behavior, according to the OU picture, 
is that the friction $\gamma$
between the impurity and the fluid  
is larger for larger temperatures.  
We will dedicate the next section to the discussion 
on the temperature dependence of $\gamma$. 

We briefly comment on the short time-lags limit ($t\lesssim10\xi/c$), 
where the measured correlator appears to decay fast and with the same slope 
for all the temperatures. 
This is particularly evident in the Log-Log plots in  
Fig.\ref{Fig:Corr}.b. 
In this regime, 
the assumptions necessary for an OU regime to be 
established are certainly not valid. 
Indeed, we are looking at timescales 
shorter than the decorrelation time of the thermal excitations
$\tauGP$, so that the collisions between 
the excitations and the impurity cannot be considered random, rapid and decorrelated as in the forcing $\mathbf{\zeta}_\mathrm{r}(t)$ in (\ref{Eq:OU}). 
It is worth noting that, for low temperatures, the velocity correlator partially recovers before the exponential decay. 
This unusual feature may be a consequence of a lack of decorrelation due to the small fraction of thermal excitations at low temperatures, which prevents the emergence of a diffusive regime. Such phenomenon requires
further investigations.

Another important prediction that can be obtained from the OU process is that the variance of the displacement $\delta_t q_i(t)=q_i(t+t_0) - q_i(t_0)$ obeys the law
\begin{equation}
\left\langle \left(\delta_t q_i\right)^2 \right\rangle = \frac{\sigma^2 \Mp}{\gamma^3}\left( \frac{\gamma }{\Mp}t - 1 + e^{-\frac{\gamma }{\Mp}t} \right).
\label{Eq:disp}
\end{equation} 
Two regimes can be identified. 
At short time-lags (but still large enough to consider the forcing $\zeta_{\mathrm{r}}(t)$ delta-correlated), 
the displacement is ballistic
\begin{equation}
\left\langle \left(\delta_t q_i\right)^2 \right\rangle \underset{t \ll M_{\mathrm{I}}/\gamma}{\longrightarrow} \frac{\sigma^2}{2\gamma \Mp}t^2.
\label{Eq:ballistic}
\end{equation} 

Conversely, after the dynamical relaxation a diffusive regime is established
\begin{equation}
\left\langle \left(\delta_t q_i\right)^2 \right\rangle \underset{t \gg M_{\mathrm{p}}/\gamma}{\longrightarrow} \frac{\sigma^2}{\gamma^2}t=2Dt,
\label{Eq:diffusive}
\end{equation} 
where we have defined the diffusion constant $D=\sigma^2/2\gamma^2$.

\begin{figure}
\includegraphics[width=.99\linewidth]{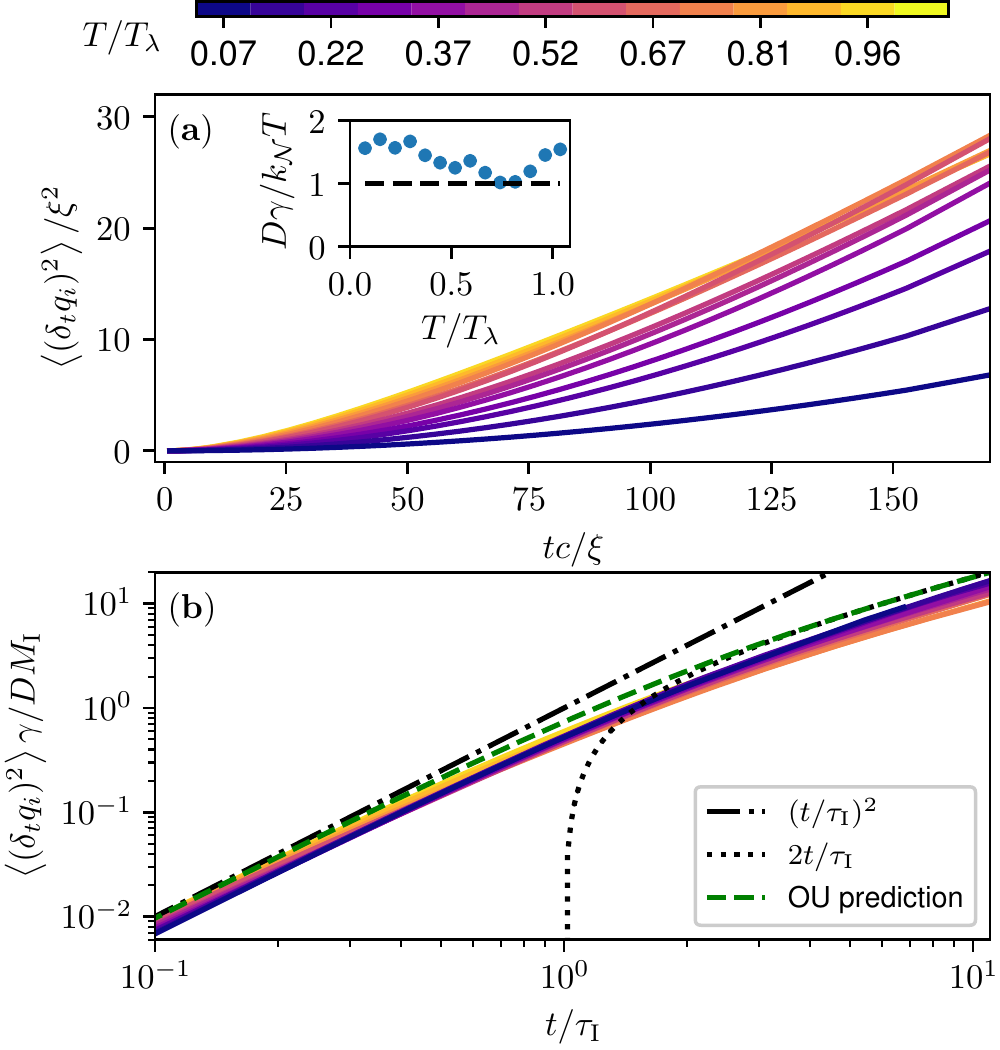}
\caption{
(\textit{Color online}) Time evolution of the averaged squared displacement for 
the impurity of size $\Rp=7.6\xi$ for different temperatures. 
Different colors are associated to different temperatures 
(same legend of Fig.\ref{Fig:PDF}). Dashed green line is 
the prediction (\ref{Eq:disp}), assuming the Einstein relation 
(\ref{Eq:EinsteinRel}), dash-dotted black line 
and dotted line are respectively the asymptotic  
(\ref{Eq:ballistic}) and (\ref{Eq:diffusive}). 
\textbf{(a)} Lin-Lin scale,  
times normalized with $\xi/c$ and distances normalized 
with $\xi$. 
\textbf{(b)} Log-Log scale, times normalized with the correlation time
$\tauB$ and distances normalized with 
the prefactor of (\ref{Eq:disp}). 
(\textit{inset}) Measured diffusion coefficient as a function 
of temperature compared with the Einstein relation (\ref{Eq:EinsteinRel}).}
\label{Fig:Disp}
\end{figure}

Finally recall that, since in the OU process we also have that $\left\langle \dot{q}_i^2 \right\rangle = \sigma^2/2\Mp\gamma=D\gamma/\Mp$, the diffusion coefficient in Eq.~\eqref{Eq:diffusive} can be related to the equipartition of energy in thermal equilibrium 
(\ref{Eq:equilibrium})  
through the Einstein relation 
\begin{equation}
D=\frac{{k_{\mathcal{N}}T}}{\gamma}. 
\label{Eq:EinsteinRel}
\end{equation}
The measurements of the average squared displacement for the impurity of size $\Rp=7.6\xi$ are shown in Fig.\ref{Fig:Disp} 
for all the temperatures analyzed, and compared with the OU predictions. 
Once the squared displacement is normalized with 
the prefactor of the prediction (\ref{Eq:disp}) and assuming 
the Einstein relation (\ref{Eq:EinsteinRel}) to estimate the diffusion coefficient,  
the separation between the ballistic regime and the 
diffusive one is apparent (bottom panel). 
The transition happens at the measured values of the dynamical correlation time $t=\tauB$, 
confirming the validity of the analysis of the 
velocity correlator. 
The diffusion coefficient $D$ is measured 
as the slope of the squared displacement 
in the diffusive regime and it is shown 
in the inset of Fig.\ref{Fig:Disp}.a. 
It is slightly larger 
than the prediction given by the Einstein relation (\ref{Eq:EinsteinRel}). 
Such trend can be the signature of a memory effect 
due to a stochastic forcing of the fluid on the impurity which is not perfectly 
delta-correlated. For instance, it 
could be traced back to the presence of coherent structures 
in the fluid or to the impurity surface fluctuations, due 
to the actual interaction between the impurity and the thermal excitations.

\subsection*{Friction modeling}

\begin{figure}
\includegraphics[width=.99\linewidth]{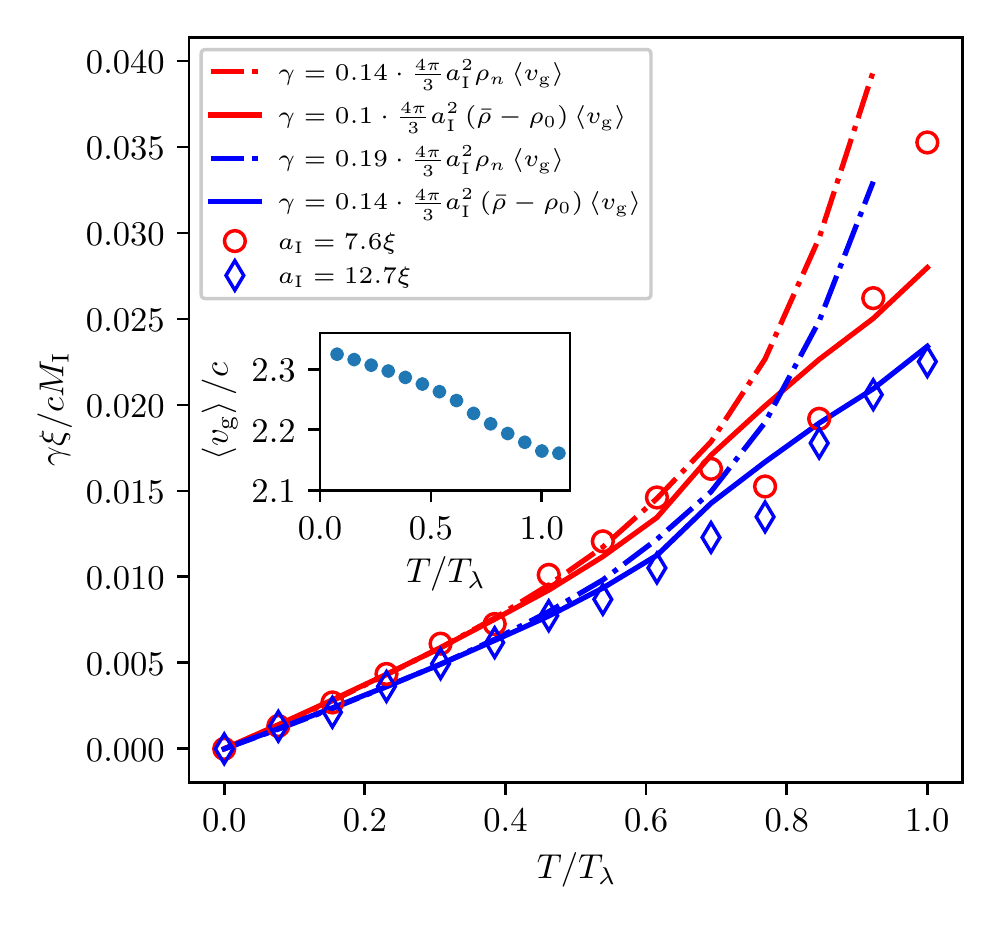}
\caption{
(\textit{Color online}) Friction coefficient $\gamma$ nondimensionalized 
by $c \Mp/\xi$ 
as a function of the temperature, 
for impurities of size $\Rp=7.6\xi$ (red circles) 
and $\Rp=12.7\xi$ (blue diamonds),
with mass density $\rho_\mathrm{I}=\bar{\rho}$. 
Dash-dotted lines are fits of the Epstein drag 
(\ref{Eq:EpsteinDrag}) using the the normal fluid density 
$\rho_\mathrm{n}$. 
Solid lines are fits of the Epstein drag using 
the density of non-condensed modes $\bar{\rho}-\rho_0$. 
(\textit{inset}) Average excitation velocity 
$\left\langle v_\mathrm{g} \right\rangle$ (\ref{Eq:vave}) 
as a function of temperature.}
\label{Fig:Gamma}
\end{figure}

In this section we show explicitly the behavior of the friction coefficient 
observed in the numerical simulations and we give a phenomenological argument 
to explain it. In Fig.\ref{Fig:Gamma}, the friction $\gamma$ is plotted as a function of 
the temperature for the two impurity sizes analyzed 
(red circles for the small one and blue diamonds for the large one).
Each value of $\gamma=M_{\mathrm{p}}/\tauB$ is estimated from 
the measured decay time $\tauB$ of the impurity velocity correlator, 
shown in the inset of Fig.\ref{Fig:Corr}.b.

In general terms, the friction $\gamma$ depends on 
the interaction between the impurity and the surrounding fluid. 
For a classical fluid there are different regimes, 
depending on the value of the Knudsen number 
$\mathrm{Kn} = \lambda_\mathrm{mfp}/\Rp$, 
where $\lambda_\mathrm{mfp}$ is the mean free path of the 
fundamental constituents of the fluid. 
If $\mathrm{Kn}\ll 1$, at the scale of the impurity, 
the fluid can be effectively considered as a 
continuous medium and the Navier--Stokes equations hold. 
As a consequence, the drag force acting on the impurity 
is the standard Stokes drag 
$\mathbf{F}_\mathrm{d}=-6\pi\Rp\eta\mathbf{\dot{q}}$ 
\cite{Batchelor}, so that the friction is related to the viscosity $\eta$ as  
\begin{equation}
\gamma = 6\pi\Rp\eta.
\label{Eq:StokesDrag}
\end{equation} 
Instead, if $\mathrm{Kn}\gg 1$, the fluid behaves as a 
dilute gas of free molecules. 
In this case, the resistance of the impurity is well described 
by the Epstein drag \cite{EpsteinDrag}: 
\begin{equation}
\mathbf{F}_\mathrm{d}=-\gamma\mathbf{\dot{q}}, \quad \gamma=\frac{4\pi}{3}C_\mathrm{d}\Rp^2\rho_\mathrm{g}\left\langle v_\mathrm{g} \right\rangle
=C_\mathrm{d} \frac{\Mp\rho_\mathrm{g}\left\langle v_\mathrm{g} \right\rangle}
{\Rp\rho_{\rm I}},
\label{Eq:EpsteinDrag}
\end{equation}
where $\rho_\mathrm{g}$ is the mass density of the gas and 
$\left\langle v_\mathrm{g} \right\rangle \gg|\mathbf{\dot{q}}|$ 
is the average velocity of the molecules. 
The pre-factor $C_\mathrm{d}$ is a dimensionless constant that depends on the interaction between the impurity and the fluid molecules. In the case of elastic collisions of the fluid excitations (specular reflection), 
a simple way of understanding the formula (\ref{Eq:EpsteinDrag}) is summarized in the following \cite{DragSimple}. If an object of mass $\Mp$ moves with velocity 
$\mathbf{\dot{q}}$ in an isotropic gas of free molecules, 
the momentum exchanged in the collision between a 
surface element $\mathrm{d}A$ and a 
molecule (assuming elastic collisions) is 
$\Delta \mathbf{p} \sim -2m_\mathrm{g}|\mathbf{\dot{q}}|\cos{\theta}\mathbf{\hat{n}}$, 
where $m_\mathrm{g}\ll \Mp$ is the molecule mass and $\theta$ is the angle between the object velocity and the outward normal 
to the surface element $\mathbf{\hat{n}}$. Assuming that the typical speed of the molecules  
$\left\langle v_\mathrm{g} \right\rangle$ is much larger than the object velocity,
the average number of collisions 
in a time interval 
$\Delta t$ is $\mathrm{d}n_\mathrm{coll} =n_\mathrm{g}\left\langle v_\mathrm{g} \right\rangle \Delta t \, \mathrm{d}A$, 
which is the number density of molecules 
$n_\mathrm{g}=\rho_\mathrm{g}/m_\mathrm{g}$ 
times the volume spanned by each molecule 
$\left\langle v_\mathrm{g} \right\rangle \Delta t \, \mathrm{d}A$. 
The infinitesimal force arising from the momentum exchange is therefore 
$\mathrm{d}\mathbf{F}_\mathrm{d} = (\Delta \mathbf{p} /\Delta t)\,\mathrm{d} n_\mathrm{coll}$. 
By symmetry, if the object is spherical, the force components orthogonal to its 
direction of motion will cancel.
Accounting for this, the net drag force results from the integration of
$|\mathrm{d}\mathbf{F}_{\mathrm{d}}|\cos\theta\left(\mathbf{\dot{q}}/|\mathbf{\dot{q}}|\right)$
over half of the sphere surface. This leads precisely to Eq. \eqref{Eq:EpsteinDrag} with $C_\mathrm{d}=1$. Considering different reflection mechanisms leads to the same equation with a different value of the pre-factor $C_\mathrm{d}$. For instance, in the case of full accomodation of the excitations with the impurity surface one gets $C_\mathrm{d}=(1+\pi/8)\sim1.39$ \cite{EpsteinDrag}.

The mean free path $\lambda_\mathrm{mfp} (T)$ in the 
FTGP model has been recently estimated in Ref. \cite{ShuklaViscosity} 
as the product of the the group velocity of the excitations 
and the nonlinear interaction time 
(i.e. the reciprocal of the spectral broadening of the dispersion relation) 
at a given temperature. 
For $\xi k_\mathrm{max}=2\pi/3$, {the value used in this work,} the mean free path $\lambda_\mathrm{mfp}$ 
turns out to lie between $10\,\xi$ and $50\,\xi$
at temperatures $T<0.7\,T_\lambda$, thus 
larger than the sizes of the 
impurities studied here   
(cfr. Fig.14 of Ref. \cite{ShuklaViscosity}). 
As a consequence, we can treat the fluid as a gas of 
free molecules and confront the measured friction with the 
Epstein drag. In particular, 
the role of ``gas molecules" in the GP fluid is 
played by the thermal excitations. 
Therefore, we can substitute the gas density 
$\rho_\mathrm{g}$ in Eq. (\ref{Eq:EpsteinDrag}) 
with the density of the non-condensed modes 
$\rho_\mathrm{g}=\bar{\rho}-\rho_0$, where $\rho_0=n_0\bar{\rho}$ 
or with the normal fluid density $\rho_\mathrm{g}=\rho_\mathrm{n}=n_\mathrm{n}\bar{\rho}$, 
computed using the momentum density correlator \cite{FosterBKT} 
(see Fig.\ref{Fig:Cond}). 
The velocity of the excitations 
$v_\mathrm{g} = \frac{\partial \omega_k}{\partial k}$ 
is averaged as: 
\begin{equation}
\left\langle v_\mathrm{g} \right\rangle = \frac {\sum_{|\mathbf{k}|\in S_\mathbf{k}}n_\mathbf{k}\frac{\partial \omega_k}{\partial k}} {\sum_{|\mathbf{k}|\in S_\mathbf{k}}n_\mathbf{k}}= \frac{ \sum_{k=1}^{k_\mathrm{max}} k^2 n_k^{1d} \frac{\partial \omega_k}{\partial k}} {\sum_{k=1}^{k_\mathrm{max}} n_k^{1d}},
\label{Eq:vave}
\end{equation}
with $n_\mathbf{k}$ the occupation number of the mode 
$\mathbf{k}\in S_\mathbf{k}=\{1\le|\mathbf{k}|\le k_\mathrm{max}\}$ 
and  $n^{1d}_k=\sum_{|\mathbf{k}|=k}n_\mathbf{k}$ its angle average.

In Fig.\ref{Fig:Gamma}, the Epstein drag prediction (\ref{Eq:EpsteinDrag}) is compared with the numerical data. 
Both using the normal fluid density (dash-dotted lines) or the density of non-condensed modes (solid lines) we get a good accordance at low temperatures, with a fitted pre-factor $C_\mathrm{d}$, whose values are of the order $0.1$. Note that in this way we are 
implicitly guessing that the impurity-excitations interaction 
is independent of temperature.  The specific values of $C_\mathrm{d}$ are reported in the legend of Fig.\ref{Fig:Gamma}. They are consistent with a reasonable scenario in which thermal waves are much less efficient in transferring momentum to the impurity with respect to the standard particles reflection mechanisms \cite{EpsteinDrag}.
We observe that $C_\mathrm{d}$ is slightly increasing with the impurity size (perhaps because of some variation of the impurity surface fluctuations) but it is independent of temperature. Note that the precise determination of radius dependence of $C_\mathrm{d}$ would require even further numerical simulations of what has been presented here.

In the inset of Fig.\ref{Fig:Gamma}, we show the temperature 
dependence of the averaged excitations velocity (\ref{Eq:vave}), which turns out to be larger than the speed of phonons because it is dominated by high wave number excitations. Note that the friction increment starts to diverge from the prediction at high temperatures. One reason is that the mean free path of the GP excitations is becoming of the same order of the impurity size and thus the viscosity starts to play a role in the momentum exchange. A second cause may be that the impurity-excitations interactions are modified because of the high nonlinearity 
of the GP waves, leading to a temperature dependence of the constant $C_\mathrm{d}$ in Eq. (\ref{Eq:EpsteinDrag}). Eventually, note that a larger discordance with the measurements at high temperature is observed if the normal fluid 
density is used. This is probably due to a lack of accuracy in the computation of $\rho_\mathrm{n}$ at high temperatures, but it also suggests that it can be more reasonable to identify the density of the excitations simply with that of the non-condensed modes.

\section{Discussion}
In this article we studied how the stochastic motion 
of an active, finite-size and immiscible impurity 
immersed in a GP quantum fluid changes when the temperature 
is varied.
We demonstrated that the interaction with the thermal 
excitations in the system always leads to 
a fast thermalization of the impurity. At time-lags 
larger than $10\xi/c$ the correlation function of 
the impurity velocity shows an exponential decay, which 
is steeper for higher temperatures. This and the impurity squared 
displacement are reminiscent of an Ornstein--Uhlenbeck
process. 

From the measurements of the velocity correlation we 
extracted the temperature dependence of the friction 
coefficient $\gamma(T)$. The clear result is that 
the impurity does not experience the typical 
Stokes drag present in a classical fluid. Indeed, 
in the case of Stokes drag, the temperature dependence of the 
friction (\ref{Eq:StokesDrag}) is through the viscosity $\eta$. 
Since the viscosity has been shown to be slightly 
decreasing with temperature in the FTGP model 
\cite{ShuklaViscosity}, it cannot explain 
the trend observed in Fig.\ref{Fig:Gamma}. 
The reason is that the settings studied are associated 
with large values of the Knudsen number, meaning that 
at the scale of the impurity 
the GP quantum fluid at finite temperature cannot be 
considered as a continuous liquid. 
On the contrary, describing phenomenologically 
the system as a gas of dilute thermal 
excitations reproduces the correct temperature 
increment of the friction $\gamma(T)$. 
Moreover, we observe a dependence of the friction with the impurity size 
compatible with the quadratic scaling $\gamma\propto\Rp^2$ predicted by the 
Epstein drag (\ref{Eq:EpsteinDrag}), despite some small deviations hidden in the prefactor $C_{\mathrm{d}}$.
In the case of Stokes drag, one should have observed a linear scaling $\gamma\propto\Rp$ that is not in agreement with our data.

We stress that the picture outlined does not apply to the 
particles typically used as probes in superfluid helium experiments 
\cite{bewley2006superfluid,LaMantiaParticles}. Indeed, besides
being liquid helium a strongly interacting system, the typical
size of those particles is $4$ orders of magnitude larger than 
the healing length. Thus, in that case 
the Knudsen number is certainly 
small enough to entail the standard Stokes drag.
However, a similar regime in terms of Knudsen number has been studied experimentally by using microspheres in liquid helium below $0.5\,K$ \cite{Schoepe}. It has been observed that the drag is determined by the ballistic scattering of quasi-particles and the temperature dependence of the friction coefficient is given by the temperature dependence of the quasi-particles density. 
Besides helium, we hope that our study may be relevant 
for future BEC experiments, in which finite-size and immiscible 
impurities can be produced in the strong repulsive regime 
of multi-component condensates 
\cite{RicaRoberts}, 
or in the study of the impurity 
dynamics in quantum fluids of light 
\cite{CarusottoLight2014,MichelSuperLight}.

A possible follow-on of the present work is the 
development of a self-consistent theory for the 
interaction between the thermal excitations and the 
impurity, which takes into account the dependence on 
the wave numbers of the colliding waves. This could give 
an analytical explanation to the small value of the prefactor $C_{\mathrm{d}}$ in Eq. 
(\ref{Eq:EpsteinDrag}) compared to the classical Epstein drag 
for elastic collisions.
Note that in a recent publication, the motion of a bright soliton moving in a thermal cloud of distinct atoms has been successfully modeled by using an OU dynamics \cite{OUsoliton}. In that case, the soliton is treated by using a wavefunction and the thermal (non-condensed) cloud as a reservoir. Although in our model the impurity is a rigid body with classical degrees of freedom, the result of \cite{OUsoliton} could inspire an analytical derivation of the OU dynamics for an impurity \eqref{Eq:OU}.
Moreover, the characterization of the motion of a 
multitude of impurities in the FTGP system can be deepened,
expanding the findings of Ref. \cite{ClusteringUmberto}.
Finally, the fundamental problem of vortex nucleation
due to fast impurities has been thoroughly 
investigated at zero temperature 
\cite{ActiveWiniecki,BrachetCritVel,FrischCritVel}, but 
few results are known in the finite temperature regime 
\cite{WinieckiNucFT,BarenghiNucFT}. In particular, 
the FTGP model coupled with impurities (\ref{Eq:HGP}) would
be a suitable framework to address the impurity-vortex 
interaction at non-zero temperature.

\acknowledgments{
The authors are grateful to Dr. D. Proment for fruitful discussions.
The authors were supported by 
Agence Nationale de la Recherche through the project GIANTE ANR-18-CE30-0020-01. GK is also supported by the EU Horizon 2020 Marie Curie project HALT and the Simons Foundation Collaboration grant Wave Turbulence (Award ID 651471).
Computations were carried out on the M\'esocentre SIGAMM hosted at the Observatoire de la C\^ote d'Azur 
and the French HPC Cluster OCCIGEN through the GENCI allocation A0042A10385.
}

%

\end{document}